\documentclass[11pt]{article}
\usepackage{moriond,epsfig,amssymb}

\def\beq{\begin{equation}}
\def\eeq{\end{equation}}
\def\bea{\begin{eqnarray}}
\def\eea{\end{eqnarray}}

\newcommand{\gevcs}      {\mbox{${\mathrm{GeV}}$}}
\newcommand{\ra}         {$\rightarrow$}

\newcommand{\clb}        {\mbox{$\mathrm{CL_b}$}}

\newcommand{\Hpm}{\mbox{$\mathrm{H}^{\pm}$}}

\newcommand{\mHpm}{\mbox{$m_{\mathrm{H}^{\pm}}$}}
\newcommand{\mHp}{\mbox{$m_{\mathrm{H}^+}$}}

\newcommand{\ee}{\mbox{${\mathrm{e}}^+ {\mathrm{e}}^-$}}

\newcommand{\bb}         {\mbox{$\mathrm{b}\bar{\mathrm{b}}$}}

\newcommand{\mH}         {\mbox{$m_{\mathrm{H}}$}}
\newcommand{\mh}         {\mbox{$m_{\mathrm{h}}$}}
\newcommand{\mA}         {\mbox{$m_{\mathrm{A}}$}}
\newcommand {\Ho}        {\mbox{$\mathrm{H}^{0}$}}
\newcommand {\Ao}        {\mbox{$\mathrm{A}^{0}$}}
\newcommand {\ho}        {\mbox{$\mathrm{h}^{0}$}}
\newcommand {\Zo}        {\mbox{$\mathrm{Z}^{0}$}}

\newcommand{\tanb}       {\mbox{$\tan\!\beta$}}

\newcommand{\Hone}{\ensuremath{\mathrm{H}_1}}
\newcommand{\Htwo}{\ensuremath{\mathrm{H}_2}}
\newcommand{\Hthree}{\ensuremath{\mathrm{H}_3}}
\newcommand{\sqrts}     {\mbox{$\sqrt{s}$}}

\begin{document}
\vspace*{4cm}
\title{Searches for the Higgs boson in Minimal Supersymmetric
CP-conserving and CP-violating Standard Model scenarios at LEP}

\author{ Pamela Ferrari }

\address{CERN, 1211 Geneve 23, Switzerland}

\maketitle\abstracts{It is important to study extended models  containing more than one physical Higgs boson 
in the spectrum. In particular, Two Higgs Doublet Models (2HDMs)~\cite{higgshunter}  
are attractive extensions of the SM, predicting new phenomena with 
the fewest new parameters. The Higgs sector in the 
Minimal Supersymmetric extension of the SM (MSSM)
is a 2HDM itself. The neutral Higgs searches performed at LEP  are showing no  
evidence of the presence of a signal and have therefore been interpreted in the context 
of 2HDMs. Depending on the model considered exclusion of  large regions of the parameter 
space can be obtained, but the existence of the lightest Higgs boson with masses lower than 90~\gevcs\ is 
not ruled out in all models by LEP.
In the MSSM at least one of the neutral Higgs bosons is predicted to
have its mass close to the electroweak energy scale; when radiative
corrections are included~\cite{radcor}, this mass should be less than about
140~\gevcs. This prediction provides a strong
motivation for searches at present and future colliders.}
\section{Introduction}

In the context of 2HDMs
the Higgs sector comprises five physical Higgs bosons: 
two neutral CP-even scalars, \ho\ and \Ho\ (with  $\mh < \mH$), one
CP-odd scalar, \Ao, and two charged scalars, \Hpm. 
Two Higgs Doublet Models are classified according to the Higgs boson couplings 
to fermions. In Type II 
models the first Higgs doublet 
couples only to down--type fermions and the 
second Higgs doublet couples only to up--type 
fermions. The Higgs sector in the 
minimal supersymmetric extension of the SM~\cite{higgshunter}
is a Type II 2HDM, in which the introduction of 
supersymmetry adds new particles and constrains the parameter space of
the Higgs sector of the model. 

At the centre-of-mass energies accessed by
LEP, the \ho, \Ho\ and \Ao\  
bosons are expected to be produced predominantly via two processes: 
the Higgsstrahlung
process \ee\ra\ho\Zo\ or  \ee\ra\Ho\Zo\
and the pair--production process \ee\ra\ho\Ao.


In the MSSM, the Higgs potential is invariant under CP transformations
at tree level.  However it is possible to explicitly or spontaneously
break CP symmetry by radiative corrections,
in particular by contributions from third generation scalar-quarks~\cite{Pilaftsis:1999qt}.   
The introduction of CP-violation is interesting since it provides a possible
solution to the cosmic baryon asymmetry~\cite{Carena:2000id}, while the
size of the CP-violating (CPV) effects occurring in the SM are far too
small to account for it.
Unlike the CP Conserving (CPC) case, in the CPV MSSM the Higgs mass eigenstates \Hone,\Htwo\ and
$\Hthree$ are not CP eigenstates. This influences predominantly
the couplings in the Higgs sector:  the mass eigenstates are mixtures
of the CP eigenstates $\mathrm{h},\mathrm{H}$ and $\mathrm{A}$ and
since only the CP-even field component couples to the $\Zo$ boson, the
individual couplings of the mass eigenstates are reduced in the CPV
with respect to the CPC case.

The size of the CP-violation scales qualitatively as the fourth power of the top mass and 
as the imaginary part of the Higgs-squark trilinear coupling~\cite{Carena:2000ks}. 

\section{Interpretation of the results}\label{sect:results}

The four LEP collaborations have searched for neutral Higgses using all LEP2 data up to the 
highest LEP energy, \sqrts $=$209~GeV. No evidence of an excess of events with 
respect to the SM expectation has been found~\cite{mssm-lep}. 
The presence of neutral Higgs bosons is tested in the MSSM as well as in a general 2HDM(II). The MSSM model 
considered is a constrained MSSM with seven parameters which are not varied independently: only a certain number of 
``benchmark sets'' are chosen where the tree level parameters $\tanb$ and $\mA$ (CPC scenario) 
or $\mHpm$ (CPV scenario) are scanned while all other parameters are fixed~\cite{benchmarks}. 

\subsection{The CPC MSSM}

\begin{figure}[p]
\centerline{
\epsfig{file=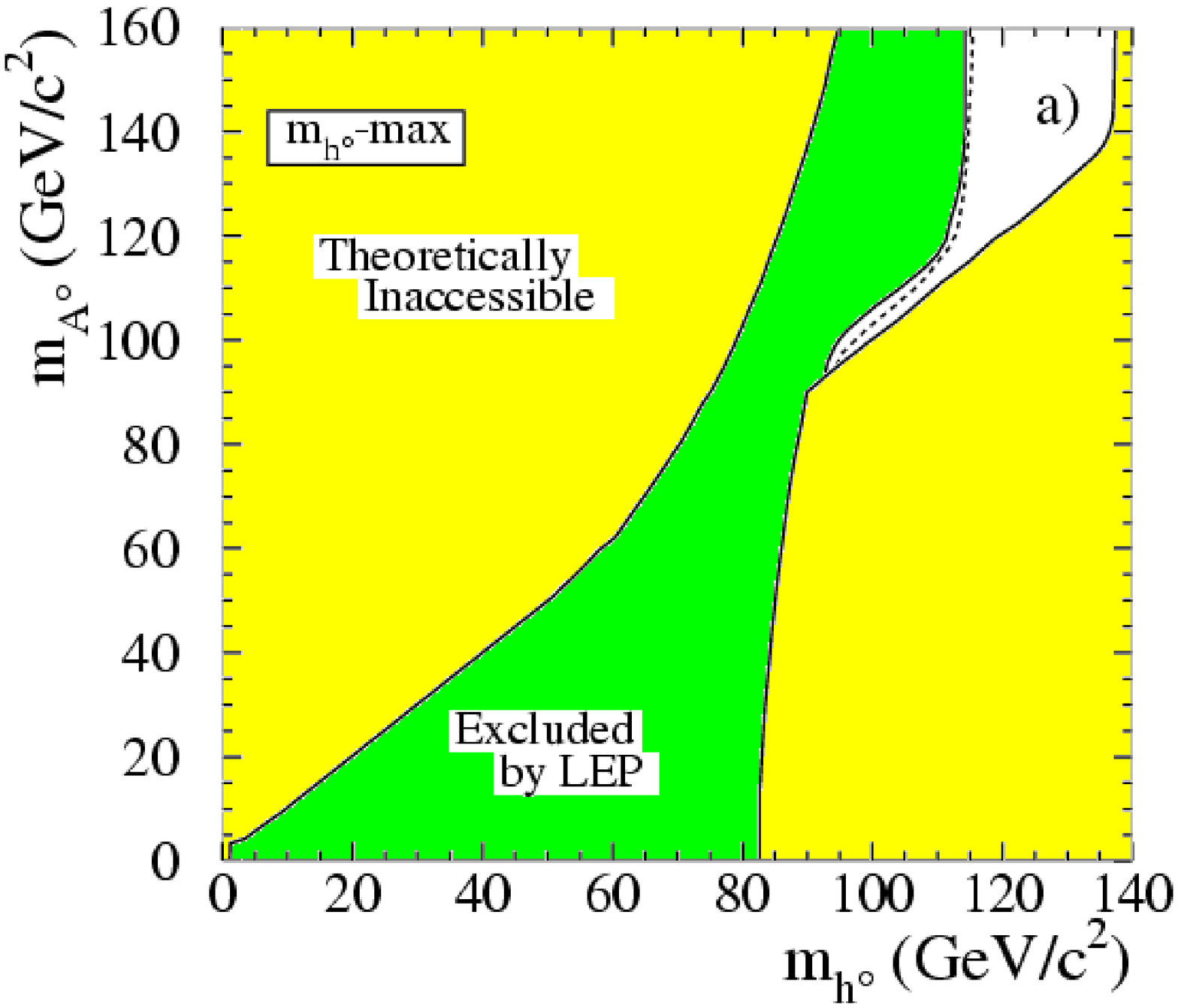,width=0.49\textwidth}
\epsfig{file=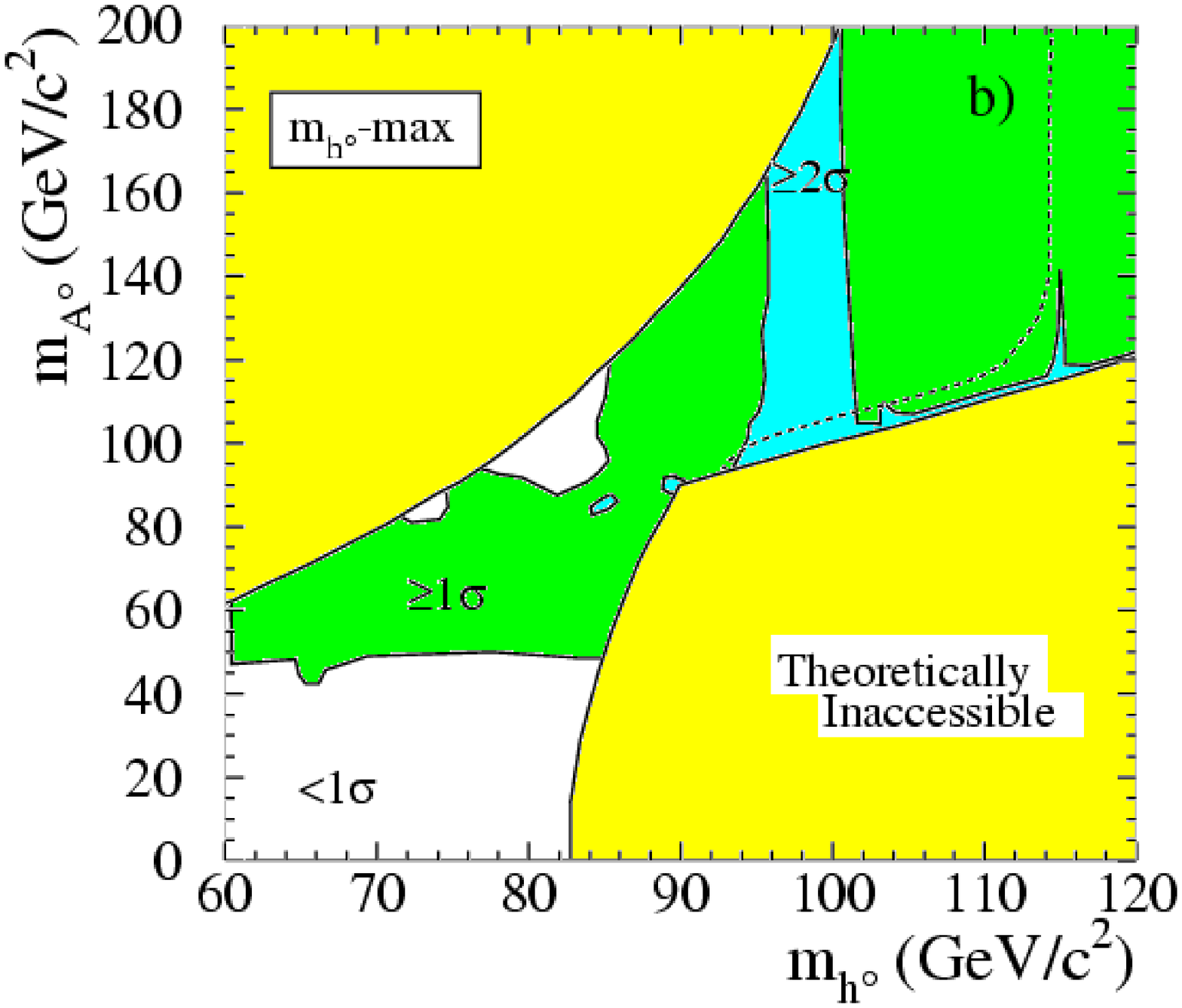,width=0.49\textwidth}
}
\caption{\label{fig:mhmax}
  a) The 95 $\%$ CL exclusion (dark grey/green area) by LEP in the (\mA,\mh) plane, for the CPC MSSM {\it \mh-max} benchmark 
scenario, with $m_{\mathrm t}=179.3$~\gevcs. The dashed lines indicate the expected exclusion on the basis of 
Monte Carlo simulations with no signal. b) Contours of the $1-CL_b$ confidence level for the {\it \mh-max} 
benchmark scenario, in the (\mh,~\mA) projection of the MSSM parameter space. The theoretical limits are indicated.
In the light-grey (blue) regions, labeled $\geq 2\sigma$, data deviate from Monte Carlo by about two to three 
standard deviations. The dashed line represents the upper edge of the region excluded at 95 $\%$ CL by this search.
}
\end{figure}

Three CPC benchmark scenarios have been studied. The {\it no-mixing} scenario where the stop mixing
parameter $X_{\mathrm{t}}$ is set to zero, giving rise to a relatively restricted MSSM 
parameter space.
The {\it \mh-max} scenario, designed to maximise the theoretical upper bound on 
\mh\ and the {\it large-$\mu$} scenario, where the $\ho$ decays to $\bb$ are suppressed, on which most of the searches are based~\cite{mssm-lep}.  
The {\it large-$\mu$} scenario is nearly completely excluded due to the contribution of the so called ``flavour-independent'' 
searches that do not depend explicitly on b-tagging  with the exception
of one thin strip at large \mA\ and large \tanb\ which is at the edge of being 
excluded at 95 $\%$ Confidence Level (CL).
Figure ~\ref{fig:mhmax}.a) shows the exclusion in the \tanb\ versus \mA\ projection in the {\it \mh-max } scenario.
The obtained lower limit on the 
lightest Higgs boson mass is of 92.9~\gevcs\ and 93.3~\gevcs\ for the {\it \mh-max} and {\it no-mixing} scenarios, respectively, 
which, taking into account the upper theoretical bound~\cite{radcor},
restricts the mass window where the lightest MSSM Higgs can be discovered 
to 93 $\lesssim$ \mh $\lesssim$ 140 \gevcs.


New benchmarks have been designed, motivated either by the fact that the planned Higgs searches at LHC may have
low sensitivity in those situations,  or by 
experimental constraints on the branching ratio of the inclusive decay
b\ra s$\gamma$ and recent measurements of the muon anomalous magnetic
moment $(g-2)_{\mu}$.  
An interpretation of the data of the four LEP experiments in those new benchmark scans is in preparation, 
but results  by the  OPAL collaboration are available~\cite{mssm-o}:
the OPAL exclusion limits obtained from the new and the traditional scans are very similar,
indicating that the LEP results for the new scans won't change significantly the conclusions 
drawn previously. 

At this point the question arises wheter by looking closely at the LEP data there is some indication of the presence 
of a Higgs. Figure~\ref{fig:mhmax}.b) shows the 1-\clb\ confidence level, which is a measure 
of the overall agreement between the data and the Monte Carlo expectation in the absence of signal. 
The two largest deviations observed, for \mh$\sim$98 \gevcs\ and \mh$\sim$115 \gevcs,
are barely exceeding two standard deviations. Recent attempts to explain those discrepancies as the production
of \ho\Zo\ and \Ho\Zo, respectively, have been made. Suitable choices of the parameters have been found
in the context of general 2HDM, CPC or CPV MSSM~\cite{excess,Carena:2000ks}, but it should be
noted that the probability of such discrepancies to be caused by a background fluctuation is quite large.

\subsection{The CPV MSSM}

The interpretation of the LEP searches  in the CPV MSSM has been performed in
the context of a  benchmark that maximises CPV~\cite{benchmarks}.
Figure~\ref{fig:CPV}.a) shows the exclusion in the \tanb\ versus \mHpm\ projection.
In this scenario no lower limit on \mh\ can be extracted, mainly because of 
the reduced couplings of the lightest Higgs to the \Zo\ boson~\cite{mssm-lep}. 
A lower bound on \tanb\ at 2.6 (2.7 expected) can be set in the assumption that 
$m_{\mathrm t} = 179.3$~\gevcs.

\subsection{The general 2HDM(II)}

\begin{figure}[p]
\centerline{
\epsfig{file=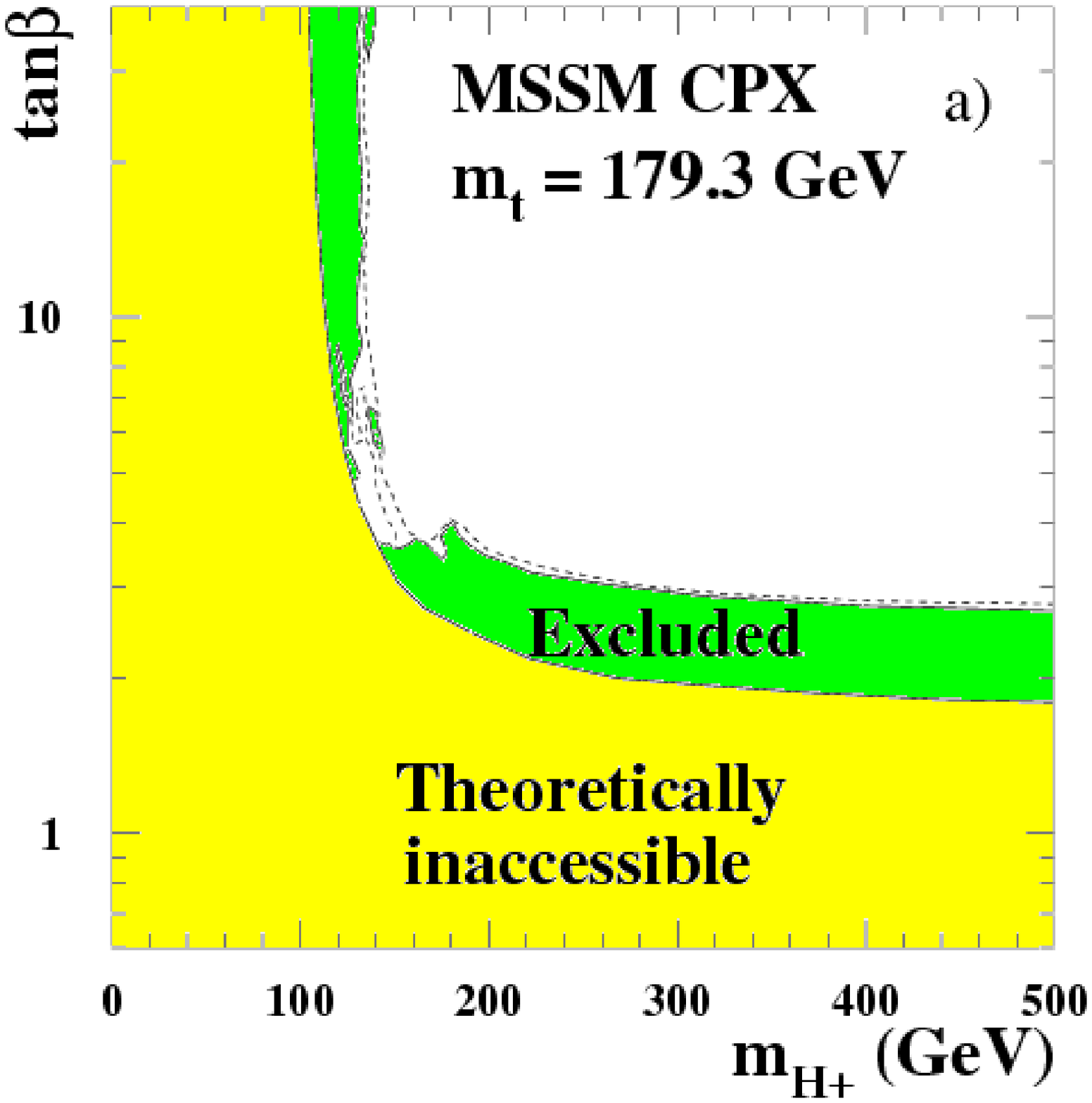,width=0.49\textwidth}
\epsfig{file=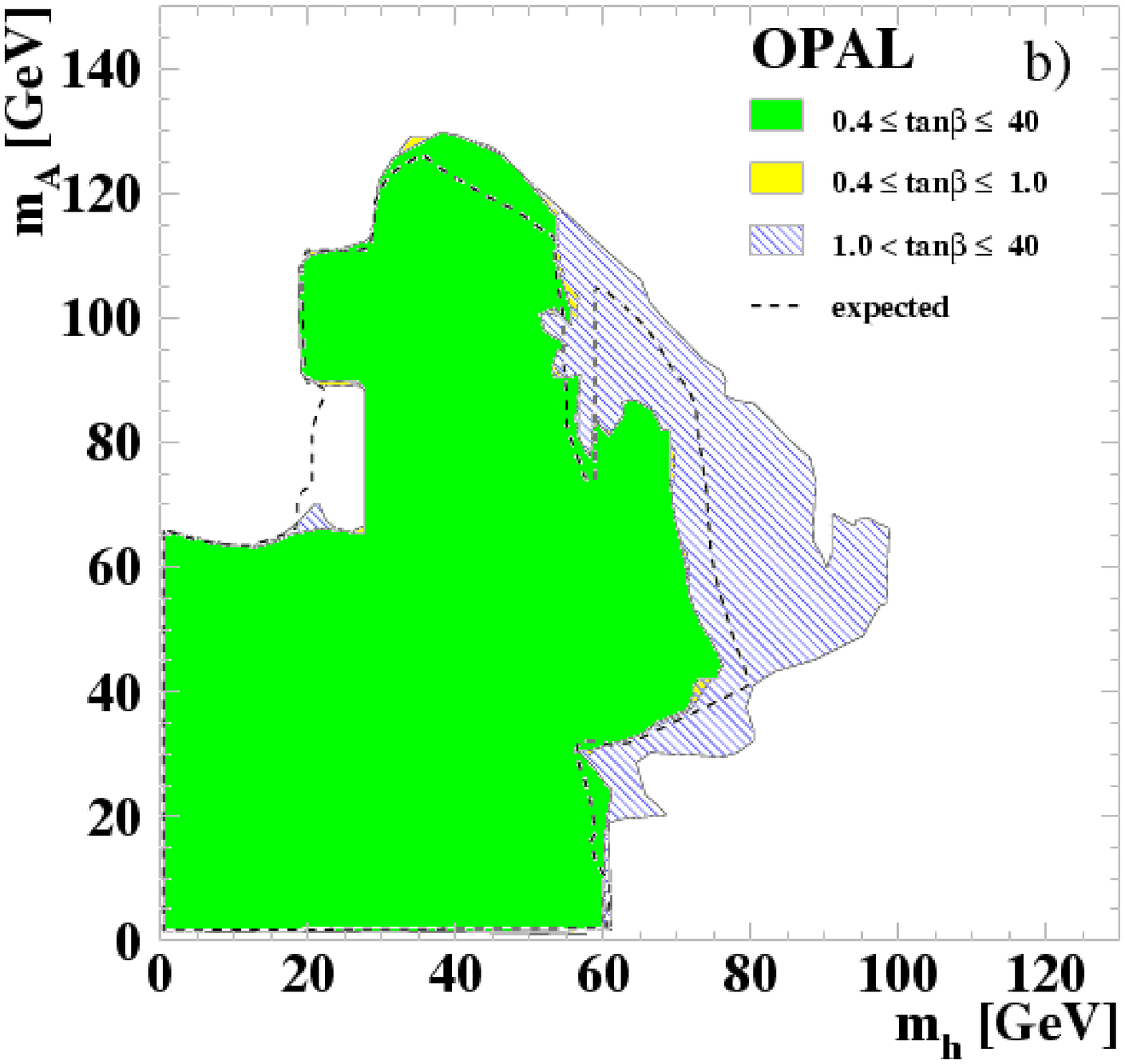,width=0.49\textwidth}
}
\caption{\label{fig:CPV} a) LEP exclusion  at 95 $\%$ CL (dark grey/green area) in the (\mHp,~\tanb) projection, 
for the CPV MSSM benchmark scenario, with $m_{\mathrm t}=179.3$~\gevcs. 
The dashed lines indicate expected exclusion in case of the absence of signal. b) Observed (dark grey/green area) 
and expected (dashed line) excluded contours in the 2HDM(II) by the OPAL collaboration in the (\mA,~\mh) projection. 
Observed exclusion for restricted \tanb\ ranges (0.4$\le$\tanb$\le$1.0 and 1.0$\le$\tanb$\le$40) are also shown.}
\end{figure}

A scan of the general 2HDM(II) has been performed by the OPAL collaboration
interpreting the same set of data used for the MSSM scans~\cite{2HDM}.
Figure~\ref{fig:CPV}.b) shows exclusion in the \mh\ versus \mA\ projection.
As in the case of the CPV MSSM it is not possible to extract a lower limit on 
\mh\ or \mA\ given the generality of the model and the fact that
the Higgs boson couplings to the \Zo, as well as the branching ratio to \bb,
can be strongly suppressed in large regions of the parameter space.

\section{Conclusions}

At LEP we have searched for Higgses in several extensions of the SM, in particular 
in the context of general 2HDM(II), CPC and CPV MSSM. No evidence of a signal 
has been found. In the CPC MSSM the mass window in which the lightest Higgs boson could 
still be found by future experiments is bounded from below at 90 \gevcs\ 
by the direct LEP searches and from above at about 140 \gevcs\ by theoretical constraints. 
In the  CPV MSSM and in general 2HDMs no lower limit on the lightest Higgs boson mass can be extracted.

\end{document}